\documentclass[oldversion]{aa}

\usepackage{amssymb}
\usepackage{times}
\usepackage{graphicx}
\usepackage[switch]{lineno}
\usepackage{xspace}
\usepackage{epsfig}
\usepackage{textcomp}
\usepackage{amsmath}
\usepackage{natbib}
\usepackage{tabularx}
\usepackage{rotating}
\usepackage{dcolumn}
\usepackage{epstopdf}

\bibpunct{(}{)}{;}{a}{}{,}

\def\gsim{\;\lower4pt\hbox{${\buildrel\displaystyle >\over\sim}$}\,}
\def\lsim{\;\lower4pt\hbox{${\buildrel\displaystyle <\over\sim}$}\,}
\begin{document}

\title{Detailed X-ray spectroscopy of the magnetar 1E$\,$2259+586}  

\subtitle{}

\author{D. Pizzocaro\inst{1}, A. Tiengo\inst{1,2,3}, S. Mereghetti\inst{1}, R. Turolla\inst{4,5}, P. Esposito\inst{1,6},  L. Stella\inst{7},  S. Zane\inst{5}, 
N. Rea\inst{6,8,9},\\ F. Coti Zelati\inst{8,9}, G. Israel\inst{7}}

\offprints{}

\institute{
  INAF-Istituto di Astrofisica Spaziale e Fisica Cosmica Milano, via A. Corti 12, 20133 Milano, Italy\label{inst1}\\
  e-mail: D. Pizzocaro, \texttt{daniele.pizzocaro@gmail.com}\and
  Scuola Universitaria Superiore IUSS, piazza della Vittoria 15, 27100 Pavia, Italy\label{inst2}\and
  INFN – Istituto Nazionale di Fisica Nucleare, Sezione di Pavia, via A. Bassi 6, 27100 Pavia, Italy\label{inst3}\and
  Department of Physics and Astronomy, University of Padova, via Marzolo 8, I-35131 Padova, Italy\label{inst4}\and
  Mullard Space Science Laboratory, University College London, Holmbury St. Mary, Surrey, RH5 6NT, UK\label{inst5}\and
  Anton Pannekoek Institute for Astronomy, University of Amsterdam, Science Park 904, Postbus 94249, 1090 GE Amsterdam, The Netherlands\label{inst6}\and
  Osservatorio Astronomico di Roma, INAF, via Frascati 33, 00078, Monteporzio Catone, Roma, Italy\label{inst7}\and
  Institute of Space Sciences (ICE, CSIC), Campus UAB, Carrer de Can Magrans s/n, E--08193 Barcelona, Spain\label{inst8}\and
  Institut d'Estudis Espacials de Catalunya (IEEC), E--08034 Barcelona, Spain\label{inst9}\\
}

  \titlerunning{Detailed X-ray spectroscopy of the magnetar 1E$\,$2259+586}
\authorrunning{D. Pizzocaro {\em et al.}}

\date{Received $<$XX-XX-2018$>$ / Accepted $<$XX-XX-2018$>$}

\abstract{
Magnetic field geometry is expected to play a fundamental role in magnetar activity. 
The discovery of a phase-variable absorption feature in the X-ray spectrum 
of SGR$\,$0418+5729, interpreted as cyclotron resonant scattering, suggests 
the presence of very strong non-dipolar components in the magnetic fields of magnetars.
We performed a deep \textit{XMM-Newton} observation of pulsar $\,$1E$\,$2259+586 to search for spectral features due to intense local magnetic fields.
In the phase-averaged X-ray spectrum, we found evidence for a broad absorption feature at very low energy ($0.7\,{\mathrm keV}$).
If the feature is intrinsic to the source, it might be due to resonant scattering and absorption by protons close to star surface.
The line energy implies a magnetic field of $\sim 10^{14}\ \mathrm G$, which is roughly similar to the spin-down measure, $\sim 6\times 10^{13} \ \mathrm G$.\\ 
Examination of the X-ray phase-energy diagram shows evidence for another absorption feature, the energy of which strongly depends on the rotational phase ($E\ga 1\ {\mathrm keV}$). 
Unlike similar features detected in other magnetar sources, notably SGR$\,$0418+5729, it is too shallow and limited to a short phase interval 
to be modeled with a narrow phase-variable cyclotron absorption line.
A detailed phase-resolved spectral analysis reveals significant phase-dependent variability in the continuum, especially above $2\,{\mathrm keV}$.
We conclude that all the variability with phase in 1E$\,$2259+586 can be attributed to changes in the continuum properties, which appear 
consistent with the predictions of  the resonant Compton scattering model.\\
}


\maketitle

\section{Introduction}

\begin{figure} 
\begin{center}
\includegraphics[width=9.3cm]{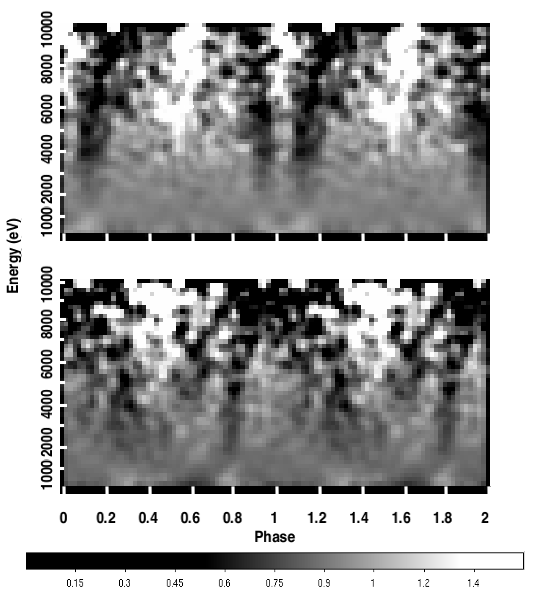}
\caption{{\em Top}: Phase-energy diagram for the EPIC PN data of observation A, obtained by binning the source counts into energy and rotational phase channels, then
normalizing to the phase-averaged energy spectrum and energy-integrated pulse profile.  
{\em Bottom}: Same for observation B, when the source was in outburst.
}
\label{fig:phaseenergy3}
\end{center}
\end{figure}

Magnetars are young isolated neutron stars characterized by an X-ray luminosity that is
often much higher than that expected from spin-down powered emission.
Historically, they have been discovered either through an analysis of their steady X-ray emission (anomalous X-ray pulsars, AXPs) or in outburst events (soft gamma repeaters, SGRs, see \citealt{2015SSRv..191..315M}). 
In the commonly adopted unified model, 
the outbursts and a significant part of the steady X-ray luminosity are due to 
the energy provided by the decay and instabilities of
very intense magnetic fields, $\gtrsim {10^{14}}\,${\rm G} (\citealt{1992ApJ...392L...9D}, \citealt{1992AcA....42..145P},
\citealt{1995MNRAS.275..255T}, \citealt{1996ApJ...473..322T}, \citealt{2002ApJ...574..332T}).\\
\indent Most AXPs and SGRs {possess} magnetic dipole fields, as inferred from their period and period derivative, which are higher than or at the high end of those of the ordinary pulsars.
However, several radio pulsars { have never been} seen to undergo bursting or flaring activity, although their
magnetic dipole fields are as strong as those of many SGRs or AXPs. Together with the recent discovery of a few SGRs with ${\rm P}$ and
$\dot{\rm P}$ indicating a dipole field well in the range of ordinary radio pulsars (\citealt{2010Sci...330..944R},
\citealt{2011ApJ...730...66L}, \citealt{2012ApJ...754...27R}, \citealt{2013ApJ...770...65R}),
this shows that a strong magnetic dipole field is not by itself a necessary nor a sufficient condition for a neutron star to be a magnetar.
Not only the field intensity, but also its topology (mainly in the stellar interior or crust) plays an important role in triggering the
magnetar activity: the strength of the toroidal component
of the internal field { is possibly} the deciding factor. This is responsible for the deformation in the neutron star crust, which imparts twists to the external magnetosphere (which results in
strong magnetospheric currents that are responsible for the nonthermal power-law component through resonant cyclotron scattering), and for crust fractures
(which are assumed to produce bursts and flares). For a review on the physics of magnetars, see, for instance, \cite{2015RPPh...78k6901T} and \cite{2017ARA&A..55..261K}.\\
\indent The discovery of a phase-variable absorption line in the X-ray spectrum of the transient low-field magnetar 
SGR$\,$0418+5729 convincingly showed that an ultra-strong component of the B field is localized in a small magnetic structure close to the stellar surface \citep{2013Natur.500..312T}. 
A somewhat similar phase-variable spectral feature was found in another transient low-field magnetar, SWIFT$\,$J1822.3-1606 \citep{2016MNRAS.456.4145R}.
The dipolar magnetic field derived from the timing properties of these two objects 
is the lowest of the currently known magnetar candidates ($\sim 6 \times 10^{12}\,{\rm G}$, \citealt{2013ApJ...770...65R}, and
$3.4(1) \times 10^{13}\,{\rm G}$, \citealt{2016MNRAS.456.4145R}, respectively). Therefore, we speculate that the detection of phase-variable absorption lines
in their spectra was probably enabled by 
the high contrast between the large-scale dipolar magnetic field and the field in small magnetic loops  that cross the line of sight only during a short rotational phase interval.\\
\indent From this point of view, a promising candidate for the search for similar phase-variable spectral features is the magnetar 1E$\,$2259+586 because its
dipole magnetic field is relatively low (B$_{\rm dip}\sim 6 \times10^{13}$ G; \citealt{2014ApJ...784...37D}).
We therefore performed a search for phase-dependent spectral features in its X-ray spectrum. 
1E$\,$2259+586 is the prototype of the old class of AXPs \citep{1981Natur.293..202F, 1983IAUS..101..445F} 
and played a significant role in the development of the unified model for magnetars.
This magnetar is a persistent X-ray emitter, with an average X-ray luminosity of $\sim 10^{35}-10^{36}\,{\rm erg/s}$ and a pulse period of $\sim 7.0\,{\rm s}$. As most magnetars, 
it also has occasional periods of bursting activity. An outburst occurred on 2002-06-18 \citep{2003ApJ...588L..93K}. {An {\em XMM-Newton} observation was performed
one week before the onset of the outburst (ID: $0038140101$, observation A in the following; see Table 1),
and another observation was made (ID: $0155350301$, B in the following) three days after the onset, while the source was still in outburst.}\\ 
\indent An efficient and quick way to 
look for phase-dependent spectral features is the visual inspection of phase-energy images,
where the photons collected from a source are binned in { energy and phase,}
and counts are normalized, so that they can be used to identify (phase-variable) spectral features.
In the phase-energy diagrams of the two {\em XMM-Newton} observations (Fig. \ref{fig:phaseenergy3}) of 1E$\,$2259+586, 
we discovered a possible time-variable absorption feature that
mainly in the observation taken in quiescence (A) resembled the feature observed in SGR$\,$0418+5729 (\citealt{2013Natur.500..312T}).
On that basis, we proposed a deep {\em XMM-Newton} observation  of 1E$\,$2259+586 to clarify the nature of this phase-dependent feature.
{In the present work we perform a detailed analysis of this observation (ID:$0744800101$, C in the following), together with that of the two archival data
sets (A and B, see Table \ref{tab:obs}). The data reduction is described in Section \ref{analysis}. In Section \ref{timing} we briefly present the timing properties 
of 1E$\,2259+586$. The spectral analysis, with special focus on the phase-resolved spectroscopy, is presented in Section \ref{results}. 
Results are discussed in Section \ref{disc}, and conclusions are drawn in Section \ref{conclusions}}.\\

\section{Data reduction}\label{analysis}
\begin{table}
\begin{center}
\caption{{ The {\em XMM-Newton} observations of 1E$\,$2259+586 analyzed in the present work}.}
\label{tab:obs}
\begin{tabularx}{\columnwidth}{ccrc}
\hline
Observation ID&Date&Duration&Count rate\\
&&& ($0.3-12\,{\mathrm keV}$)\\
\hline
0038140101&2002-06-11&$52\,{\rm ks}$&$9.72 \pm 0.019$ cts/s\\ 
0155350301&2002-06-21&$31\,{\rm ks}$&$18.77 \pm 0.034$ cts/s\\ 
0744800101&2014-07-29&$112\,{\rm ks}$&$10.62 \pm 0.006$ cts/s\\ 
\hline
\end{tabularx}
\end{center}
\end{table}
{We compare our observation C with the two deepest archival observations (A and B, see \citealt{2004ApJ...605..378W} for details on these observations)}.
Observation A (2002-06-11) had a duration of $52\,{\rm ks}$, with a net exposure time for the EPIC\footnote{EPIC ({\em European Photon Imaging Camera}) 
is located at the focus of the three grazing-incidence multi-mirror X-ray telescopes
that constitute the main instrument of {\em XMM-Newton}; it consists of three CCD cameras: one PN and two MOS (\citealt{2001A&A...365L..18S}, \citealt{2001A&A...365L..27T}).}
PN camera of $24.9\,{\rm ks}$. 
Observation B (2002-06-21) had a duration of $31\,{\rm ks}$, with a net exposure time for the EPIC PN camera of $18.5\,{\rm ks}$. Observation B was performed while the source was 
 in outburst, {at a flux level that was about { three} times higher than the quiescent flux}.
Observation C (2014-07-29)  had a duration of  $112\,${\rm ks},
with a net exposure time for the EPIC 
PN camera of $100\,{\rm ks}$.\\  
\indent We {focused on} the EPIC data from the PN instrument because of its higher time resolution
($5.7\,$ms, Small Window mode, whereas the MOS cameras in Small Window mode have a time resolution of $0.3\,$s).
In order to characterize possible phase-dependent features also in the low-energy region, we selected the $0.3-12.0\,{\mathrm keV}$ energy band, which is broader than 
the energy range commonly adopted in this type of studies (e.g., \citealt{2008ApJ...686..520Z}).
{The source extraction region is a circle with a radius of $40\,''$. For the background, we chose
two rectangular regions ($90\,''\times90\,''$ and $90\,''\times60\,''$) at the border of the $\sim 4\,' \times 4\,'$ PN Small Window, 
in order to minimize the contribution from the point spread function (PSF) of the central source.}

\begin{figure} 
\begin{center}
\includegraphics[width=9.3cm]{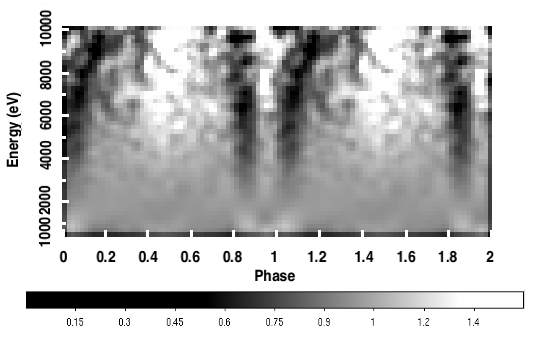}
\caption{Phase-energy diagram for the EPIC PN data of observation C, normalized 
as in Fig. \ref{fig:phaseenergy3}. The V-shaped absorption profile is visible between phase $0.8$ and $1.2$. 
}
\label{fig:phaseenergy4}
\end{center}
\end{figure}

\section{Timing analysis}\label{timing}
\indent From the EPIC PN data set of observation C we generated a barycentered event file using the \texttt{barycenter} SAS tool. 
From the barycentered time series, we calculated the rotation period with a folding analysis ($P=6.979164(1)\,{\rm s}$).
{We also performed phase-connection analysis. The two values are consistent within $1\, \sigma$. The uncertainties were evaluated through Monte Carlo simulations.
With this value for the rotation period, we calculated}
the phase for the barycentered events with the \texttt{phasecalc} tool.
We produced a phase-energy image for the EPIC PN observation by binning the source counts  
into energy and rotational phase channels (phase bin width: $0.02$; energy bin width: $200\,{\mathrm eV}$). This image was then
normalized to the phase-averaged energy spectrum  and the energy-integrated pulse profile. 
{The visual inspection of such a phase-energy diagram
(Fig. \ref{fig:phaseenergy4})} shows a possible phase-dependent structure, as suggested (with a poorer counting statistics) by the archival observation A, namely
a V-shaped feature spanning the plotted energy range. 
The data {reduction and the creation} of the phase-energy images performed on the X-ray data of observation C were also applied to
the archival observations of 1E$\,$2259+586 (A and B, Fig. \ref{fig:phaseenergy3}).\\
\begin{figure} 
\begin{center}
\includegraphics[width=9.3cm]{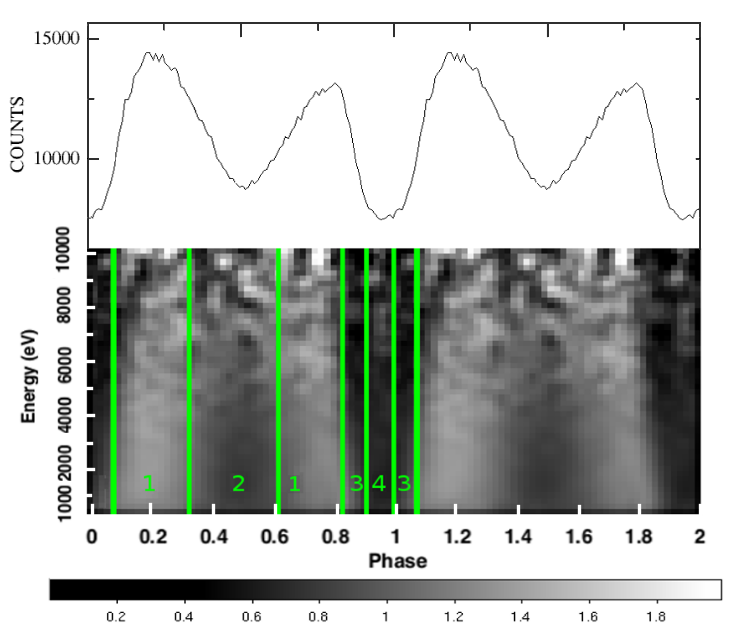}
\caption{{\em Top}: EPIC PN light curve in the energy range $0.3-12.0\,{\mathrm keV}$ for observation C. The double-peak pulse profile is clearly visible.
{\em Bottom}: Phase-energy diagram for the EPIC PN data obtained by binning the source counts into energy and rotational phase channels, then
normalizing to the phase-averaged energy spectrum for the same observation (same as in Fig. \ref{fig:phaseenergy4}, but not normalized to the energy-integrated pulse profile). 
The four phase bins used for the phase-resolved spectroscopy on the EPIC spectrum are identified by the green bars and numbers.  
}
\label{fig:phaseenergy}
\end{center}
\end{figure}
\begin{figure} 
\begin{center}
\includegraphics[width=8.8cm]{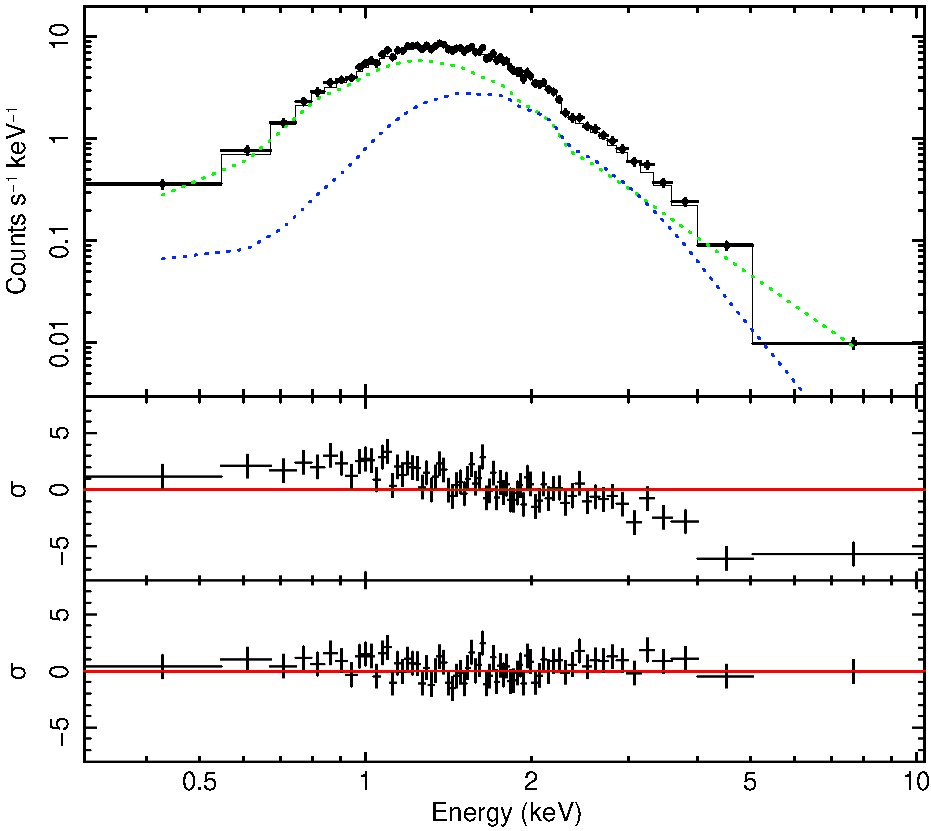}
\caption{{ Example of a spectral fit in one of the $50$ phase bins we used for the phase-resolved spectroscopy. This is one of the bins located inside the bars of the V-shaped 
feature}. {\em Upper panel}: Results of the best fit for the \texttt{const*cyclabs*gabs*TBabs*(pegpwrlw + bbodyrad)} model (best-fit model for the phase-averaged spectrum plus 
a cyclotron absorption line) to the EPIC PN spectrum. The blackbody (blue line) and power-law (green line) components are plotted. 
{\em Lower panels}: Residuals (in standard deviation units) for the best fit obtained without ($\chi^2$=539.2 for 347 d.o.f., null hypothesis probability $\simeq10^{-10}$),
and with ($\chi^{2}=360.2$ for $344$ d.o.f., null hypothesis 
probability $=0.26$) the cyclotron 
absorption line. The best-fit parameters for the cyclotron absorption line are $E=1.0\,{\mathrm keV}$ and width $W=4.0\,{\mathrm keV}$. 
A factor $5$ graphic rebin is used. 
}
\label{fig:badfit}
\end{center}
\end{figure}

\section{Spectral analysis}\label{results}
\indent {We generated the spectrum of observation C for the PN, and also for the two MOS, as a cross-check. We performed both a phase-averaged 
and a phase-resolved spectral analysis 
by dividing the observation into phase bins and fitting different spectral models in the XSPEC \citep{1996ASPC..101...17A} spectral analysis package. 
We compared the spectrum of observation C with the archival observations A and B.}
\subsection{Phase-averaged spectroscopy}
In the XSPEC spectral analysis package, we adopted a phenomenological model made by a power-law
and a blackbody component, modified by photoelectric absorption {(\texttt{TBabs*(pegpwrlw + bbodyrad)})}. When we fit this model to the phase-averaged PN spectrum of observation C,
no acceptable fit was obtained (null hypothesis probability $\sim 10^{-10}$). 
A good fit was instead obtained by {including} a Gaussian absorption component (\texttt{gabs}).
The best fit ($\chi^2$=1203.3 for $1157$ degrees of freedom, null hypothesis probability$=0.18$) is obtained for a broad ($\sigma \simeq 3\,{\mathrm keV}$) absorption line 
centered at $\sim 0.7\,{\mathrm keV}$. The results of the best fit are reported in Table \ref{tab:res} and Fig. \ref{fig:averaged}.
{Owing to the large number of counts in each spectral bin, systematic errors may be more relevant than the statistical errors. We therefore tested the 
possibility that this absorption feature was consistent with the systematic uncertainties. We plotted the residuals of the best fit
obtained without the \texttt{gabs} component in units of data/model ratio. In the energy range around $\sim 0.7\,{\mathrm keV}$, the residuals are $\text{about } 10\%$ or higher, 
which is  far higher than the dynamical range of systematic uncertainties observed in EPIC-PN spectra ($\lesssim 4\%$, see, e.g., \citealt{2014A&A...564A..75R} and references therein). 
This means that the observed absorption feature cannot be explained in 
terms of an incorrect calibration of the spectral response.}\\
\indent We analyzed the data from the two MOS cameras as well by fitting the
spectral model adopted for the PN to the two MOS spectra simultaneously.
A broad Gaussian absorption line at low energy is required here as well to obtain a satisfactory fit ($\chi^{2}=2083.51$, 
$1949$ d.o.f., null hypothesis probability = $1.7\cdot10^{-2}$; without line: $\chi^{2}=2696.19$; $1951$ d.o.f., null hypothesis probability = $\sim 10^{-27}$,
Fig. \ref{fig:averaged_mos}).
The best-fit values for the parameters of the Gaussian absorption feature { (\texttt{gabs} LineE$=0.87\pm0.02\,{\mathrm keV}$, \texttt{gabs} Sigma$=0.20\pm0.01\,{\mathrm keV}$) 
and some other parameters from the MOS are slightly different 
from those of the PN.} 
A simultaneous fit to the MOS and PN spectra, leaving only
a normalization factor for the different cameras free to vary, is not acceptable. 
Nonetheless, compatibility was obtained by considering a 
systematic error of $\sim 1.5-2\%$ in the model parameters, which is within the range of the cross-calibration uncertainties between the three EPIC cameras 
(e.g., \citealt{2014A&A...564A..75R}).
{A similar broad absorption feature at $E \simeq 0.65\,{\mathrm keV}$ is also  evident in the low-energy spectrum of observation A, but not in observation B,
when 1E$\,$2259+586 was much brighter and displayed a different continuum spectrum.\\
\begin{figure} [!htb]
\begin{center}
\includegraphics[width=9.0cm]{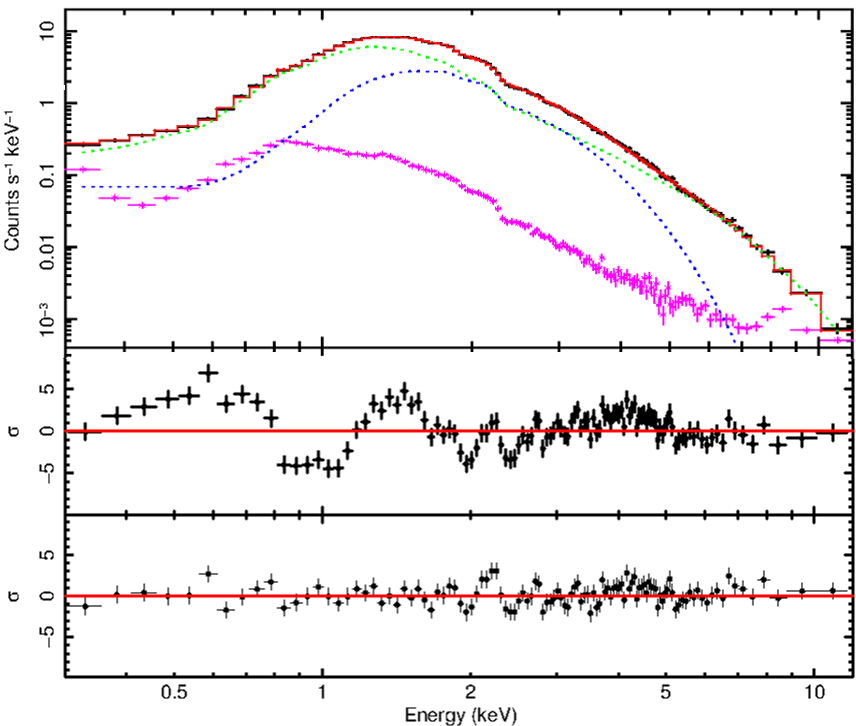}
\caption{{\em Upper panel}: Results of the best fit for the EPIC PN phase-averaged spectrum of observation C (red line) obtained with a 
\texttt{gabs*TBabs*(pegpwrlw + bbodyrad)} model.
The blackbody (blue line) and power-law (green line) components are plotted{, as well as the spectrum of the background (magenta).}
{\em Lower panels}: Residuals (in standard deviation units) for the best fit obtained without (top) and with (bottom) the \texttt{gabs} component.
A factor $10$ graphic rebin is used.}
\label{fig:averaged}
\end{center}
\end{figure}
\begin{figure} [!htb]
\begin{center}
\includegraphics[width=9.0cm]{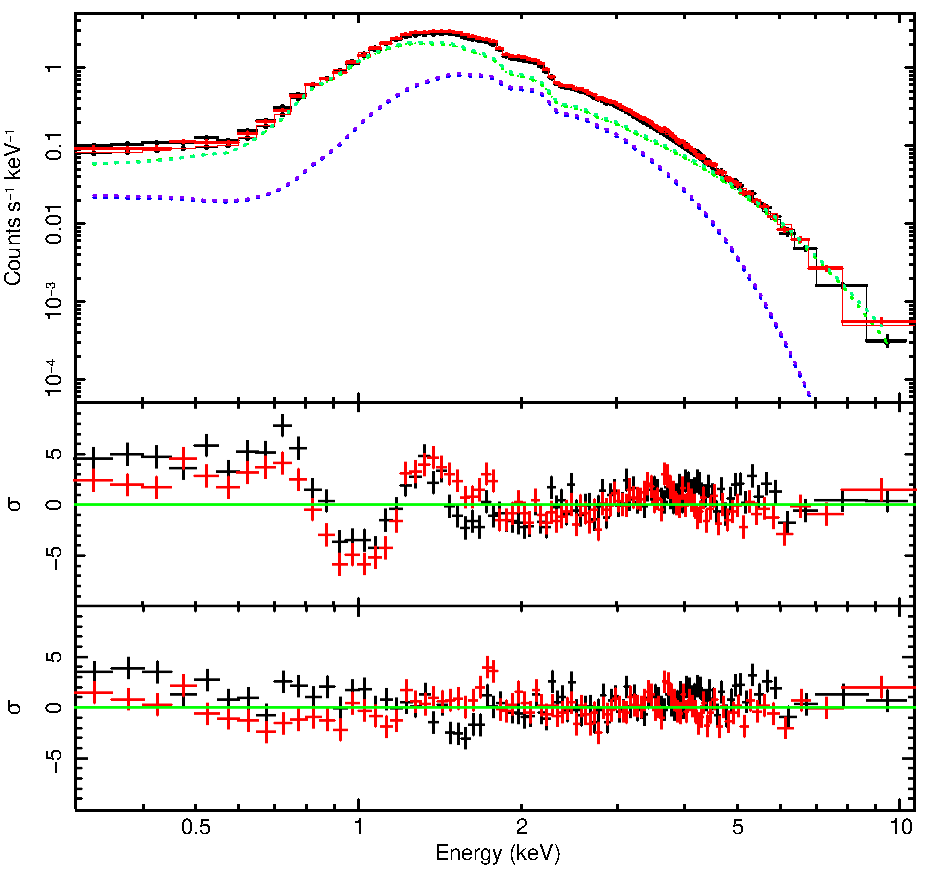}
\caption{{\em Upper panel}: Results of the best fit for the EPIC MOS1 (black) and MOS2 (red) phase-averaged spectrum of observation C obtained with a 
\texttt{gabs*TBabs*(pegpwrlw + bbodyrad)} model.
The blackbody (blue and magenta line) and power-law (green and yellow line) components are plotted.
{\em Lower panels}: Residuals (in standard deviation units) for the best fit obtained without (top) and with (bottom) the \texttt{gabs} component.
A factor $10$ graphic rebin is used.}
\label{fig:averaged_mos}
\end{center}
\end{figure}

\subsection{Phase-resolved spectroscopy}
To characterize the high-energy phase-dependent spectral features identified through visual inspection of the phase-energy diagram,
we {first} followed the procedure adopted by \cite{2013Natur.500..312T} {to detect the possible signature of a narrow phase-dependent cyclotron absorption line.} 
Taking advantage of the high counting statistics, 
we divided the source PN event list of observation C into $50$ phase bins of equal width ($0.02$){.
We} fit the model for the entire observation (\texttt{gabs*TBabs*(pegpwrlw + bbodyrad)}) to the spectrum in each phase bin, with all parameters frozen to their best-fit 
value, {including} a free multiplicative factor (\texttt{const}) to account for different count rates in different bins. 
The spectra of the phase bins that are associated with the V-shaped dark feature that is apparent in the phase-energy image were inconsistent with the continuum emission model. 
The spectrum of one of the phase bins where a poorü fit (null hypothesis probability $<0.003$) was obtained is shown in Fig. \ref {fig:badfit}.
We {therefore} tried to fit the same model plus a cyclotron absorption line (\texttt{const*cyclabs*gabs*TBabs*(pegpwrlw + bbodyrad)}).
The best fit was obtained for a very broad absorption line at energies above $1\,{\mathrm keV}$.
{However, as suggested by the residuals in the top panel in Fig. \ref{fig:badfit}, a fit with equivalent quality can be 
obtained by substituting the cyclotron line component with a high-energy cutoff or by letting the parameters of the power-law component free to vary.}
This indicates that some kind of phase-dependent spectral variability is present, 
but it cannot be straightforwardly modeled with a narrow absorption feature as in the case of SGR$\,$0418+5729 \citep{2013Natur.500..312T}.\\
\begin{table*}
\scriptsize
\begin{center}
\caption{Best-fit model parameters (errors at $1\,\sigma$ confidence level) for the phase-averaged spectrum and the spectra of the four phase bins.
{Linked parameters are reported only for bin 1.}}
\label{tab:res}
\begin{tabularx}{\textwidth}{lcccccccccr}
\hline
Bin&Factor&PL Photon Index&PL flux&gabs LineE&gabs Sigma&gabs Strength\footnotemark[1]&${\rm N_{H}}$&bbodyrad kT&bbodyrad radius\footnotemark[2]\\
&&&$10^{-12}\,{\rm erg\,/\,cm^{2} s}$&${\mathrm keV}$&${\mathrm keV}$&&$10^{22}\,{\rm cm^{-2}}$&${\mathrm keV}$&km\\
\hline\\
Avg.&-&$3.86\pm 0.03$&$10.0\pm 0.2$&$0.68\pm0.07$&$0.33\pm0.03$&$0.46\pm0.15$&$1.21\pm0.04$&$0.449\pm0.003$&$2.36\pm 0.06$\\
1&$1.00$&$3.86\pm 0.03$&$11.7\pm 0.2$&$0.79\pm0.05$&$0.27\pm0.03$&$0.27\pm0.08$&$1.24\pm0.03$&$0.442\pm0.003$&$2.77\pm 0.08$\\
2&$0.63$&$3.71\pm 0.03$&$15.7\pm 0.6$&''&''&''&''&''&''\\ 
3&$0.67$&$4.13\pm 0.04$&$9.3\pm 0.6$&''&''&''&''&''&''\\ 
4&$0.56$&$3.80\pm 0.04$&$11.4\pm 0.7$&''&''&''&''&''&''\\ 
\\
\hline
\end{tabularx}
\tablefoottext{1}{Line depth} 
\tablefoottext{2}{Assuming a distance for 1E$\,$2259+586 of $3.2\,{\rm kpc}$ \citep{2012ApJ...746L...4K}}
\end{center}
\end{table*}
\indent Because the phase-resolved spectroscopy described above did not provide conclusive results on the phase variability in the high-energy ($\gtrsim 5\,{\mathrm keV}$)
region, we decided to divide the observation into broader phase intervals to better highlight the spectral evolution in phase. 
The X-ray light curve of 1E$\,$2259+586 shows a double peak (Fig. \ref{fig:phaseenergy}, upper panel). 
The phase-energy image (not normalized for the pulse profile) displayed in the lower panel of Fig. \ref{fig:phaseenergy} shows that the apparent V-shaped feature is located within      the absolute minimum of the pulsation profile. 
In the high-energy region ($\gsim 5\,{\mathrm keV}$) three distinct phase regions can be distinguished: two with a relative depletion of events (bars of the V) and a 
central region with a relative excess of events. At lower energy, the low signal-to-noise ratio of the data 
prevents us from clearly observing the tip of the V that connects the two bars.
We divided the observation into the four phase bins shown in Fig. \ref{fig:phaseenergy}, which according to our inspection 
of the pulse profile and residual of the continuum modeling maximize the spectral differences along the phase axis.
These phase intervals correspond to the two maxima ($0.05-0.32$ plus $0.62-0.83$, bin$\,1$);
the relative minimum of the pulsation ($0.32-0.62$, bin\,2); the event-depleted bars of the V
on the sides of the absolute minimum of the pulsation ($0.83-0.88$ plus $0.98-0.05$, bin\,3);
and the phase region located in the trough of the V, corresponding to the absolute minimum ($0.88-0.98$, bin\,4)\footnote{For { bins 1 and 3}, which are defined as the sum of two non-adjacent phase intervals, we have then checked that the spectra of the sub-bins are consistent with each other.}.\\
\indent We simultaneously fit the phenomenological model used for the phase-averaged spectrum (\texttt{gabs*TBabs(pegpwrlw+bbodyrad)})
to the {spectra of the four bins. 
We included a free overall normalization factor (\texttt{const}) to account for the different flux levels of the spectra in different bins. 
We linked the value of the parameters across the $\text{four}$ bins, leaving only some of them free to vary from bin to bin.
By leaving only the overall normalization (\texttt{const}) free to vary across the four bins, no acceptable fit is obtained ($\chi^2$=4127.1 for $3192$ d.o.f., 
null-hypothesis probability $\sim 10^{-27}$).  In the same way, linking the parameters of the power-law component and leaving 
the parameters of the blackbody component free to vary, no acceptable fit is obtained. On the other hand, by leaving the parameters 
of the power-law component free to vary independently (with linked blackbody parameters), 
we obtained a very good fit ($\chi^2$=3221.8 for $3128$ d.o.f., null-hypothesis probability $=0.118$). 
The spectra of the four bins together with the best-fit model are shown in Fig. \ref{fig:spectra},
and the corresponding best-fit values for the 
photon index and normalization for the power-law component are reported in Table \ref{tab:res}.\\
\indent
For each of the four spectra, we computed a contour plot of the best-fit photon index against the power-law normalization (Fig. \ref{fig:cont})}. 
The parameters of the blackbody component were the same for each bin.
{Because we kept the blackbody normalization constant across the four bins (which was made possible by introducing an overall normalization factor,
\texttt{const})
the value of the power-law component normalization  in each bin represents the relative flux of the power-law component with respect to the blackbody.
This allowed us to evaluate the relative strength of the power-law and blackbody components
in the spectrum.}
From the contour plot, it is evident that {regardless of the overall normalization factor}, { all spectra are inconsistent with 
each other, except for bins$\,$1 and 4}.\\
\begin{figure} [!htb]
\begin{center}
\includegraphics[width=9.2cm]{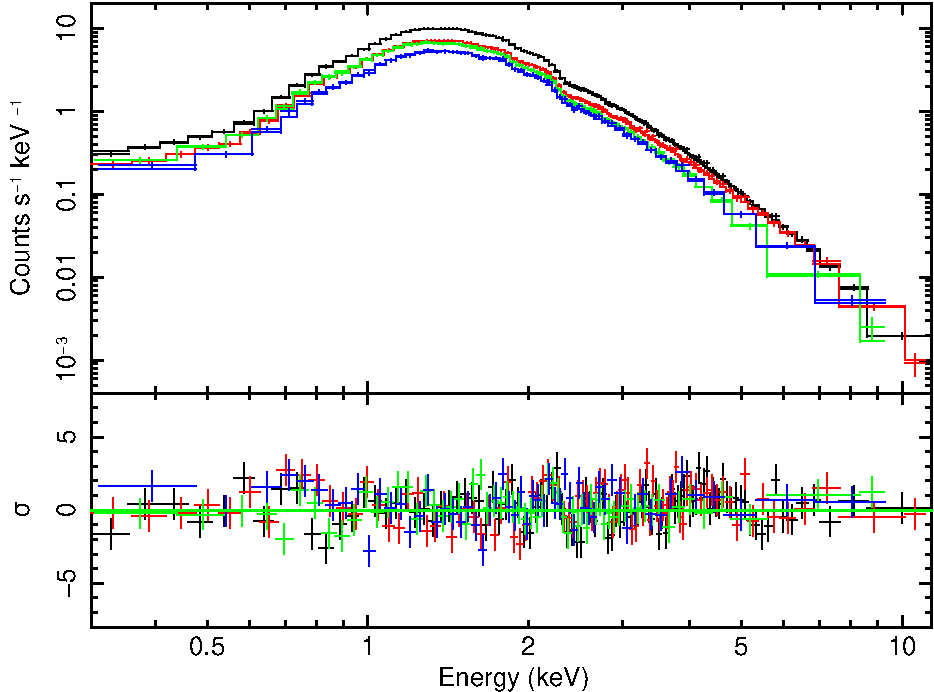}
\caption{Results of the joint fit for the four phase-resolved EPIC PN spectra of observation C. Black: Bin 1.
Red: Bin 2. Green: Bin 3. Blue: Bin 4. The lower panel shows the residuals (in standard deviation units) for each spectrum. A factor $10$ graphic rebin is used.
}
\label{fig:spectra}
\end{center}
\end{figure}
\indent Following \cite{2013Natur.500..312T}, we tested the addition of a cyclotron absorption line component to 
the phase-averaged model and fit it to the spectra of the $\text{four}$ bins,
leaving only the parameters of the line free to vary {across different bins, and linking all the others}. 
{ As for the spectra of some of the narrow phase intervals, the addition of a (broad) cyclotron absorption line to the
model was sufficient to obtain an acceptable fit also for all the spectra of the four
broad phase intervals.}
The results we obtained are comparable in terms of the goodness of fit to what we obtained by
leaving the parameters of the power-law component free to vary in the simpler continuum model we adopted for the phase-averaged spectrum.
There is therefore no statistical evidence that a cyclotron absorption line is needed to describe the phase variability observed in the phase-energy diagram of 1E$\,2259+586$.
\\
\indent As shown by \cite{2007ApJ...660..615F} and \cite{2008MNRAS.386.1527N}, spectra produced by resonant Compton scattering (RCS) of thermal photons by magnetospheric currents
show a distinct phase-dependence that affects the power-law tail at energies above $\sim 5\,{\mathrm keV}$. 
Motivated by this, we performed a fit to the phase-averaged
spectrum with the RCS model (\texttt{ntznoang} in XSPEC; \citealt{2008MNRAS.386.1527N}; \citealt{2009MNRAS.398.1403Z}). The best fit ($\chi_r^2=1.21$ for $1183$ d.o.f.)
yields $T=0.38\,{\mathrm keV}$, $\beta_{bulk}=0.20$, $\Delta\phi= 0.86\,$rad and an overall normalization factor of $0.56$ for the model parameters 
(here $T$, $\beta_{bulk}$ , and $\Delta\phi$ are the surface temperature, the bulk electron velocity in units of $c,$ and the twist angle, respectively; see 
\citealt{2008MNRAS.386.1527N} for more details). We computed the phase-resolved spectra for the model parameters obtained from the fit of the phase-averaged spectrum, 
and a few combinations of $\chi$ and $\xi$, {the two angles that the star rotation axis makes with the
line of sight and the dipole axis}. While the phase-averaged spectrum is quite insensitive to 
the source geometry, phase-resolved spectra depend on $\chi$ and $\xi$, {so that different choices of $\chi$ and $\xi$ yield different phase-dependent variability in the
continuum spectrum.} Results confirm that the observed spectral variations with phase can indeed be recovered within the RCS model for
suitable geometries.\\
\indent {We considered the possibility that { also} the $0.7\,{\mathrm keV}$ feature discovered in the phase-averaged spectrum shows a phase dependence.} 
The distribution of residuals in both the 50-bin and 4-bin analysis does not provide any evidence of phase variability for the $0.7\,{\mathrm keV}$ feature. 
Nonetheless, Fig. \ref{fig:phaseenergy4} shows possible structures in the
$E < 1\,{\mathrm keV}$ region: an apparent depletion of events at phase $\sim 0.2$ and $\sim 0.8$. 
However, also { when we divided} the observation into two phase bins ($0.12-0.24$ plus $0.74-0.82$ and
$0.24-0.74$ plus $0.82-0.12$) and fit the phase-averaged
model to their spectra, leaving the \texttt{gabs} energy free to vary, we did not find significant variability.\\

\section{Discussion}\label{disc}

Absorption features in the X-ray spectra of neutron stars are considered an astrophysical ``holy grail" because by comparing the observed and intrinsic energy 
of lines of unambiguous origin, the gravitational redshift can be measured.
This depends on the neutron star compactness ($M/R$), { which would allow placing} constraints on the equation of state (EOS) of nuclear matter. Furthermore, 
understanding the physical origin of spectral features is fundamental  for determining the composition of a possible atmosphere and thus fit the thermal emission spectra 
with the most appropriate models. 
\begin{figure} 
\begin{center}
\label{fig:cont}
\includegraphics[width=9.0cm]{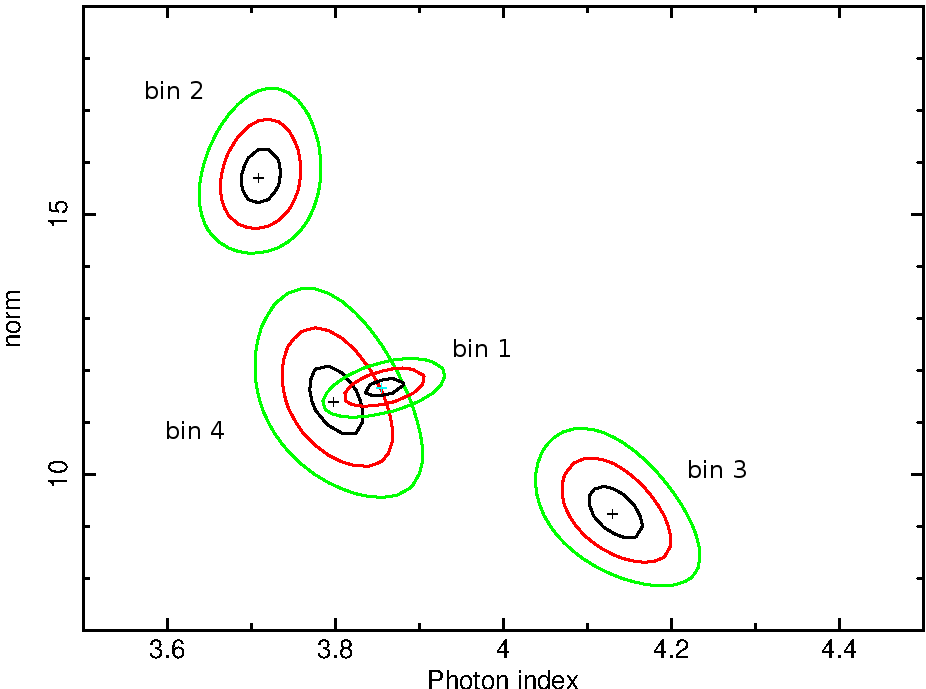}
\caption{Contour plots for the $1, 2,\text{and }3\,\sigma$ contours of the values of photon index and normalization (unabsorbed flux in the $2-10\,{\mathrm keV}$ band in units of $10^{-12}\,{\rm erg\,s^{-1}\,cm^{-2}}$
{divided by the constant factors reported in Table \ref{tab:res}}) 
of the power-law component of the best-fit model (\texttt{const*gabs*TBabs*(pegpwrlw+bbodyrad)})
for the four phase bins used in the phase-resolved spectroscopy.
}
\label{fig:cont}
\end{center}
\end{figure}
Broad absorption features centered at some hundreds of ${\mathrm eV}$ have repeatedly been observed  in the spectra of most X-ray dim isolated neutron stars (XDINSs), as 
single or harmonically spaced features, which often display spectral phase variability (\citealt{2003A&A...403L..19H}, \citealt{2004A&A...424..635H}, \citealt{2004ApJ...608..432V}, 
\citealt{2005ApJ...627..397Z}, \citealt{2007Ap&SS.308..619S}, \citealt{2007Ap&SS.308..181H}, \citealt{2015ApJ...807L..20B}, \citealt{2017MNRAS.468.2975B}). 
XDINSs are a class of  radio-quiet isolated neutron stars that { are} characterized by purely thermal X-ray spectra.
The apparent lack of contamination from nonthermal magnetospheric emission in their spectra  makes these stars very good candidates for spectral { studies that are aimed} at 
constraining the EOS.
Phase-dependent harmonically spaced absorption lines  were   revealed
in the radio-quiet pulsar 1E$\,$1207.4-5209  (fundamental at $0.7\,{\mathrm keV}$, plus two or possibly three harmonics,  \citealt{2002ApJ...574L..61S}, \citealt{2002ApJ...581.1280M},
\citealt{2003Natur.423..725B}, \citealt{2004A&A...418..625D}), 
as well as in 
the rotating radio transient 
PSR$\,$J$1819-1458$  (at $\sim 0.5\,{\mathrm keV}$ and $\sim 1\,{\mathrm keV}$, \citealt{2007ApJ...670.1307M}). 
More recently, absorption lines have also been found in standard rotation-powered radio pulsars (e.g., PSR  J1740+1000, \citealt{2012Sci...337..946K};  PSR B0628--28,
\citealt{2018A&A...615A..73R}, PSR J0656+1414, \citealt{2018arXiv181011814A}).\\
\indent In order to obtain a good fit for the high statistics   spectrum   of 1E$\,$2259+586 provided by our deep EPIC exposure, we had to  introduce a broad Gaussian
absorption line ($\sigma\sim 0.3\,{\mathrm keV}$) at $\sim 0.7\,{\mathrm keV}$. 
This line was detected with  high significance in the phase-averaged spectrum.
The properties of this broad feature are similar to those of the lines detected in some XDINS, and when we interpret it as a  cyclotron line, it implies a magnetic 
field of $6\times10^{10}$ (1+z) G  or  1.1$\times10^{14}$ (1+z) G in the case of electrons or protons, respectively. 
This is to be compared with the dipole field of $\sim 6\times10^{13}\,$G derived from the timing parameters of 1E$\,$2259+586. 
Using a typical value for the neutron star $M/R$ ($0.12\,M_{\odot}/{\rm km}$, {\citealt{2019JPhG...46b4001S}}) and the following gravitational redshift ($z=0.25$), 
we obtain a value in the proton scattering scenario that is consistent, within an order of magnitude with the magnetic field inferred from timing. 
The large width and the lack of variability of this line throughout the magnetar rotation are consistent with proton cyclotron scattering in the global 
magnetic field close to the surface of the neutron star. In particular, following \cite{2001ApJ...560..384Z}, we derive an expected line width of $\sim 0.35\,{\mathrm keV}$,
which is in very good agreement with our best-fit value of $0.33\,{\mathrm keV}$. 
However, given the apparent absence of  phase variability of this feature, we cannot completely rule out the possibility of a spurious origin that is due to calibration issue.
{On the other hand, the detection of this absorption feature in the EPIC MOS spectra as well, although with slightly different best-fit 
parameters, suggests that it is not a calibration artifact.}
Another possibility is that the feature we found in 1E$\,$2259+586 is due to an inadequate modeling of the interstellar absorption in our spectra. 
\\
\indent Our long observation of 1E$\,$2259+586 was primarily motivated by the features observed in the phase-energy images derived from previous {\em XMM-Newton} observations
of this source (see Fig. \ref{fig:phaseenergy3} and Fig. \ref{fig:phaseenergy4}). They resemble those found in the transient magnetars SGR$\,$0418+5729 and SWIFT$\,$J1822.3-1606, which were  
interpreted as proton cyclotron lines in localized regions of strong magnetic field  (\citealt{2013Natur.500..312T}; 
\citealt{2016MNRAS.456.4145R}). 
The phase-energy images displayed in Figs. \ref{fig:phaseenergy3} and \ref{fig:phaseenergy4} are quite similar to those of SGR$\,$0418+5729 and SWIFT J1822.3-1606.
In particular, the behavior
of 1E$\,$2259+586 in outburst (observation B, bottom panel of Fig. \ref{fig:phaseenergy3}), with the  single slightly inclined dark feature discovered in the present work 
(at phase $\sim 0.8$) resembles 
that of SWIFT$\,$J1822.3-1606, whereas the V-shaped dark feature that appears during the quiescent observations (A, C) {recalls} the feature that was observed in SGR$\,$0418+5729.
{However, the V-shaped feature in our observation C looks somewhat different from the feature observed in SGR$\,$0418+5729 because 
its relative strength is lower, especially at the lowest energies, and it appears less extended in rotational phase.}\\
\indent If a proton cyclotron
interpretation were acceptable, then this variation would suggest a different geometry of the strong multipolar magnetic field components and/or of the hot 
regions that generate the photons that interact with the magnetospheric protons between the quiescence and outburst phase. 
However, the spectral variability of 1E$\,$2259+586 
cannot be straightforwardly modeled with a proton cyclotron model as in the basic model proposed for SGR$\,$0418+5729 \citep{2013Natur.500..312T}.
We therefore described the high-quality phase-resolved spectrum obtained from our observation C with a phenomenological
model, where the bulk of the X-ray emission is provided by the sum of a blackbody and a power-law component. 
In this context, the spectral changes of  1E$\,$2259+586 along its pulsation phase can be fully attributed to a 
phase-dependence of the power-law slope and 
of its intensity with respect to that of the blackbody component. 
{The results of the fit of the \texttt{ntzang} model to the spectra of the $\text{four}$ broad phase bins suggest that the phase variability in the continuum of the spectrum of 
observation C is indeed consistent with RCS in the magnetosphere.}   
It is interesting to note that despite
their significantly different flux, the spectral shapes of the two maxima and of the absolute minimum are consistent with each other.
On the other hand, the hardest and softest spectra are observed in the secondary minimum and in two narrow-phase intervals around 
the absolute minimum.
\\

\section{Conclusions}\label{conclusions}

We have carried out a long  {\em XMM-Newton} observation of the persistent magnetar 1E$\,$2259+586, with the main objective to  search for phase-dependent spectral features.
We found that the phase-averaged spectrum is well fit by a phenomenological model consisting of the combination of an absorbed blackbody plus power law with a 
Gaussian absorption line ($E\sim 0.7\,{\mathrm keV}$) accounting for a phase-independent deficit of counts at low energies.
This broad ($\sim 0.3\,{\mathrm keV}$) absorption feature resembles the features that are detected in the X-ray spectra of other isolated neutron stars, and in particular, in most XDINSs.
If this line is caused by proton cyclotron resonant scattering, the inferred magnetic field is not much different from that derived from pulsar 
spin-down parameters. 
On the other hand, if it is due to electrons, the derived field is more than two orders of magnitude smaller than the surface dipole field, implying an origin high in 
the stellar magnetosphere.\\
\indent We found  that the spectrum of 1E$\,$2259+586 varies significantly as a function of the rotational phase, especially at energies $>2\,{\mathrm keV}$.  
Although the phase-energy  images of 1E$\,$2259+586 in quiescence and in outburst resemble those of the low field magnetars SGR$\,$0418+5729 and SWIFT$\,$J1822.3-1606, 
respectively, 
in this case, the phase-resolved X-ray spectra cannot be fit by a simple model with a narrow phase-variable cyclotron absorption feature. 
We found instead that the phase-dependent spectral variations can be entirely explained
in terms of changes in the continuum emission, namely a significant variation ($>3\sigma$) in the hardness and normalization of 
the power-law spectral component. We suggest that this variability can be attributed to RCS in the twisted global dipole field of the magnetar.\\


\begin{acknowledgements}
This work is based on
observations obtained with {\em XMM-Newton}, an ESA science mission with
instruments and contributions directly funded by ESA Member States and
NASA. PE acknowledges funding in the framework of the project ULTraS, ASI--INAF contract N.\,2017-14-H.0.
LS acknowledges financial contributions from ASI-INAF agreements 2017-14-H.O and  I/037/12/0 and
from iPeska research grant (P.I. Andrea Possenti) funded under the INAF call PRIN-SKA/CTA (resolution
70/2016). FCZ is supported by grant AYA2015-71042-P.

\end{acknowledgements}

\bibliography{j2259_final_20180911}
\bibliographystyle{aa}

\end{document}